\documentclass{cjpsuppl}
\usepackage[english]{babel}
\usepackage[dvips]{graphicx}
\usepackage{cite}

\newcommand{\lsim}{\,{\buildrel < \over {_\sim}}\,}
\newcommand{\gsim}{\,{\buildrel > \over {_\sim}}\,}
\newcommand{\sqrtsNN}{\sqrt{s_{\scriptscriptstyle{{\rm NN}}}}}
\newcommand{\av}[1]{\left\langle #1 \right\rangle}

\newcommand{\gev}{\mathrm{GeV}}
\newcommand{\tev}{\mathrm{TeV}}
\newcommand{\fm}{\mathrm{fm}}

\newcommand{\cm}{\mathrm{cm}}

\newcommand{\mum}{\mathrm{\mu m}}

\newcommand{\PbPb}{\mbox{Pb--Pb}}

\newcommand{\NN}{\mbox{nucleon--nucleon}}

\renewcommand{\AA}{\mbox{nucleus--nucleus}}
\newcommand{\RAA}{R_{\rm AA}}

\newcommand{\pt}{p_{\rm t}}
\renewcommand{\d}{{\rm d}}

\newcommand{\Dz}{\mbox{$\mathrm {D^0}$}}

\newcommand{\Jpsi} {\mbox{J\kern-0.05em /\kern-0.05em$\psi$}\xspace}

\begin{document}

\title{Charm production and energy loss at the LHC with ALICE}
\authori{A. Dainese}
\addressi{Universit\`a degli Studi di Padova and INFN, via Marzolo 8, 35131 Padova, Italy
\\~\\{\rm for the ALICE Collaboration}
}
\authorii{}    \addressii{}
\authoriii{}   \addressiii{}
\authoriv{}    \addressiv{}
\authorv{}     \addressv{}
\authorvi{}    \addressvi{}
\headtitle{Charm production and energy loss at the LHC with ALICE}
\headauthor{A. Dainese}
\lastevenhead{A. Dainese: Charm production and energy loss at the LHC with ALICE}
\refnum{}
\daterec{29 September 2004;\\final version 29 September 2004}
\suppl{A}  \year{2004} \setcounter{page}{1}
\maketitle

\begin{abstract}
The latest results on the ALICE performance for production 
and in-medium QCD energy loss measurements of charm particles at the LHC
are presented.
\end{abstract}

\section{Introduction}
\label{intro}

The ALICE experiment~\cite{alice} will study proton--proton (pp), 
proton--nucleus (pA) and nucleus--nucleus (AA) collisions at the LHC, 
with centre-of-mass energies per nucleon--nucleon (NN) pair, $\sqrtsNN$,
from $5.5~\tev$ (for \PbPb) to $14~\tev$ (for pp). 

ALICE is the dedicated heavy-ion experiment at the LHC and its
primary physics goal is the investigation of 
the properties of QCD matter at the energy densities of several hundred 
times the density of atomic nuclei that will be reached in central 
\PbPb~collisions. In these conditions
a deconfined state of quarks and gluons, the Quark--Gluon Plasma (QGP), 
is expected to be formed~\cite{muller}.
As we shall detail in the following Section~\ref{probes}, 
heavy quarks, and hard partons in general, probe this medium
via the mechanism of QCD energy loss~\cite{muller,armesto}. 
An estimate of the medium-induced 
suppression of open charm mesons is presented in 
Section~\ref{Dquenching}, along with the expected ALICE sensitivity 
for the measurement of this effect.

The unique features of the ALICE detector, 
such as the low-momentum acceptance
and the excellent particle identification, will also allow a rich program 
of pp physics, complementary to the those 
of ATLAS, CMS and LHCb. One outstanding 
example is the measurement of the charm production cross section, which
is described in Section~\ref{results}.

\section{Heavy-quark production and energy loss}
\label{probes}

Heavy quarks are produced in primary partonic scatterings with large 
virtuality $Q$ (momentum transfer) and, thus, 
on short temporal and spatial scales, $\Delta\tau\sim \Delta r\sim 1/Q$. 
In fact, the minimum virtuality $Q_{\rm min}=2\,m_{\rm Q}$ 
in the production of a $\rm Q\overline Q$ pair implies a space-time scale 
of $\sim 1/(2\,m_{\rm Q})\simeq 1/2.4~\gev^{-1}\simeq 0.1~\fm$ (for charm). 
Therefore, in \AA~collisions, 
the hard production process itself should not be affected by the successive 
formation of the high-density deconfined medium. 

Given the large virtualities that characterize the production of heavy quarks,
the cross sections in \NN~collisions can be calculated 
in the framework of collinear factorization and perturbative QCD (pQCD). 
The inclusive differential $\rm Q\overline{Q}$ cross section is written as:
\begin{eqnarray}
 \d\sigma_{{\rm NN\to Q\overline{Q}}X}(\sqrtsNN,m_{\rm Q},\mu_R^2,\mu_F^2) 
&=&
   \sum_{i,j=q,\overline q,g} 
   f_i(x_1,\mu_F^2)\otimes f_j(x_2,\mu_F^2)\otimes   \nonumber \\
&& \d\hat\sigma_{ij\to {\rm Q \overline Q}\{k\}}
(\alpha_s(\mu_R^2),\mu_F^2,m_{\rm Q},x_1,x_2),
\label{sigQQ}
\end{eqnarray}
where $\d\hat \sigma_{ij\to {\rm Q \overline Q}\{k\} }$ is the perturbative
partonic hard part, calculable as a power series in the strong
coupling $\alpha_s(\mu_R^2)$, which depends on the renormalization scale 
$\mu_R$;
currently, calculations are performed up to next-to-leading order (NLO),
$\mathcal{O}(\alpha_s^3)$. 
The nucleon Parton Distribution Functions (PDFs) 
for each parton $i(j)$ at momentum fraction $x_1 (x_2)$
and factorization scale $\mu_F$, which can be interpreted as the virtuality 
of the hard process, are denoted by $f_i(x,\mu_F^2)$.
At LHC energies, there are large uncertainties, of about a factor 2,
on the charm and beauty production cross sections, estimated by varying the 
values of the masses and of the scales $\mu_F$ and 
$\mu_R$ (much smaller uncertainties, $\approx 20\%$, 
arise from the indetermination in the PDFs)~\cite{yrhvq,notehvq}. 
Different predictions for 
the D-meson cross section as a function of the 
transverse momentum ($\pt$) will be shown in 
Section~\ref{results}, along with the ALICE capability to constrain the 
pQCD parameter space for charm production.

For hard processes such as heavy-quark production, in the absence of nuclear 
and medium effects, a \AA~collision 
would behave as a superposition of independent NN collisions. 
The hard processes yields would then scale from pp to AA 
proportionally to the number of inelastic NN collisions (binary scaling).
Applying binary scaling to `average' 
pQCD results and taking into account nuclear 
shadowing effects ---the PDF suppression at small $x$ in the nucleus---
we expect about 115 $\rm c\overline{c}$ and about 5 $\rm b\overline{b}$ 
pairs per central (5\% $\sigma^{\rm tot}$) 
\PbPb~collision at $\sqrtsNN=5.5~\tev$ (these numbers are obtained at NLO
using the HVQMNR program~\cite{hvqmnr} 
with $m_{\rm c}=1.2~\gev$ and $\mu_F=\mu_R=2\,\mu_0$
for charm and $m_{\rm b}=4.75~\gev$ and $\mu_F=\mu_R=\,\mu_0$ for beauty, 
where $\mu_0^2\equiv m_{\rm Q}^2+(p_{\rm t,Q}^2+p_{\rm t,\overline Q}^2)/2$;
the PDF set is CTEQ 4M~\cite{cteq4} corrected for nuclear shadowing 
according to the EKS98 parameterization~\cite{eks}).

Deviations from binary scaling correspond to deviations from unity of the 
{\it nuclear modification factor} (here defined for D mesons):
\begin{equation}
\label{eq:RAAD}
  R_{\rm AA}^{\rm D}(\pt)\equiv \frac{1}{{\rm binary~NN~collisions}} \times
    \frac{{\rm d}N^{\rm \scriptscriptstyle D}_{\rm AA}/{\rm d}\pt}
       {{\rm d}N^{\rm \scriptscriptstyle D}_{\rm pp}/{\rm d}\pt}.
\end{equation}
Note that at the LHC, pp and \PbPb~collisions will be run at different energies
and an extrapolation of the measured pp yields will have to be applied to 
define $\RAA$. Such extrapolation can be reliably done by means of pQCD, 
since it was shown~\cite{yrhvq} that the ratios of calculation results at 
different $\sqrt{s}$ are basically insensitive to the choice of masses and 
scales.

Experiments at the Relativistic Heavy-Ion Collider (RHIC) 
have shown that the nuclear modification factor is a 
powerful tool for the study of the interaction of the produced hard partons 
with the medium formed in the collision. The suppression of a factor 4--5 
of $\RAA$ for charged hadrons and neutral pions for $\pt\gsim 5~\gev/c$
observed in central Au--Au collisions at $\sqrtsNN=130$--$200~\gev$ 
is interpreted as a consequence of parton energy loss in a dense 
medium~\cite{harris}.

An intense theoretical activity has developed around the subject of parton 
energy loss via medium-induced gluon 
radiation~\cite{gyulassywang,bdmps,wiedemann,glv}.
In our sensitivity studies (see Section~\ref{Dquenching}) 
we have considered the BDMPS model in the multiple soft 
scattering approximation~\cite{bdmps}. 
Its main features are summarized in the functional form 
of the average energy loss for a high-energy ($E\to\infty$) 
hard parton with path length $L$ in the medium: 
\begin{equation}
\label{eq:avdE}
\av{\Delta E} \propto \alpha_s\,C_R\,\hat{q}\,L^2.
\end{equation}
$C_R$ is the Casimir coupling factor (3 if the considered hard 
parton is a gluon, 4/3 if it is a quark); $\hat{q}$ is the transport 
coefficient of the medium, defined as the average transverse momentum 
squared transferred to the projectile per unit mean free path and it is 
proportional to the density of scattering centres (gluons) in the medium and 
to the typical momenta exchanged in interactions with such centres; the 
characteristic $L^2$-dependence arises from the non-Abelian nature of QCD.  
The average energy loss is independent of the initial parton energy $E$,
if this is very large. In the more realistic case of parton energies
of $\sim 10$--$50~\gev$, there is an intrinsic dependence of the radiated 
energy on the initial energy, determined by the fact that the former cannot 
be larger than the latter, $\Delta E\leq E$. A rigorous theoretical 
treatment of this finite-energy constraint is at present lacking in the 
BDMPS framework and approximations have to be adopted, which introduce 
uncertainties in the results~\cite{pqm}.

Due to the large values of their masses the charm and beauty quarks are 
qualitatively different probes from light partons. 
Heavy quarks with momenta up to $40$--$50~\gev/c$ propagate with 
a velocity  significantly lower than the velocity of light. 
As a consequence
gluon radiation at angles $\Theta$ smaller than their mass-to-energy ratio 
$\Theta_0=m_Q/E_Q$ is suppressed by destructive
interference~\cite{dokshitzerdeadcone}. 
The relatively depopulated cone around the heavy-quark direction with 
$\Theta<\Theta_0$ is called `dead cone'.
In Ref.~\cite{dokshitzerkharzeev}, on the basis of an 
approximation of the dead-cone effect, charm quarks were predicted to 
lose much less energy than light quarks. 
A recent detailed calculation~\cite{armestomassive} confirms this 
qualitative feature, although the effect is found to be quantitatively
smaller than in Ref.~\cite{dokshitzerkharzeev}.

At the LHC, the abundant production of charm quarks will allow
to study the mass dependence of parton quenching
and, thus, to test experimentally these effects.

\section{Measurement of open charm production with ALICE}
\label{results}

One of the most promising channels for open charm detection is the 
$\rm D^0 \to K^-\pi^+$ decay (and its charge conjugate), which 
has a branching ratio ($BR$) of about $3.8\%$.
The expected production yields per unit of rapidity, $y$, at central rapidity 
for $\rm D^0$ (and $\rm \overline{D^0}$) mesons decaying in a 
$\rm K^\mp\pi^\pm$ pair,
estimated~\cite{notehvq} on the basis of NLO pQCD calculations, 
are $BR\times\d N/\d y=5.3\times 10^{-1}$ in central ($5\%~\sigma^{\rm tot}$) 
Pb--Pb collisions at $\sqrtsNN=5.5~{\rm TeV}$ and 
$BR\times\d N/\d y=7.5\times 10^{-4}$ in pp collisions at 
$\sqrt{s}=14~{\rm TeV}$.
\begin{figure}[!t]
  \begin{center}
  \includegraphics[width=0.8\textwidth]{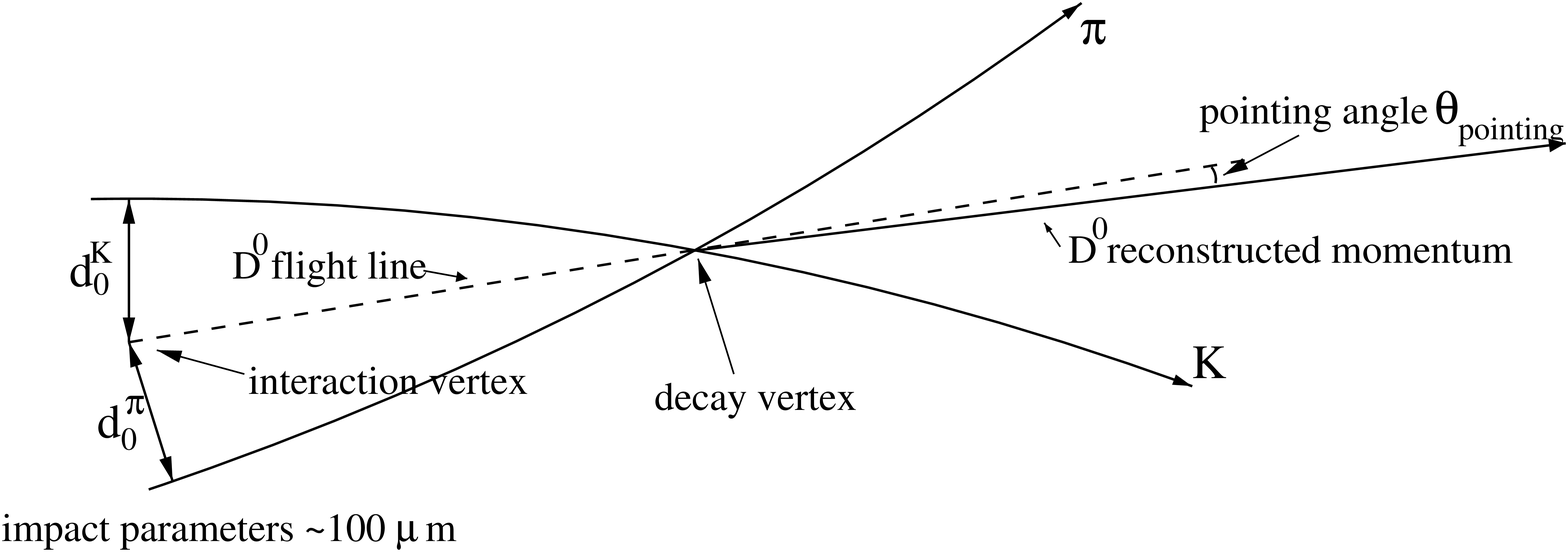}
  \includegraphics[width=0.4\textwidth]{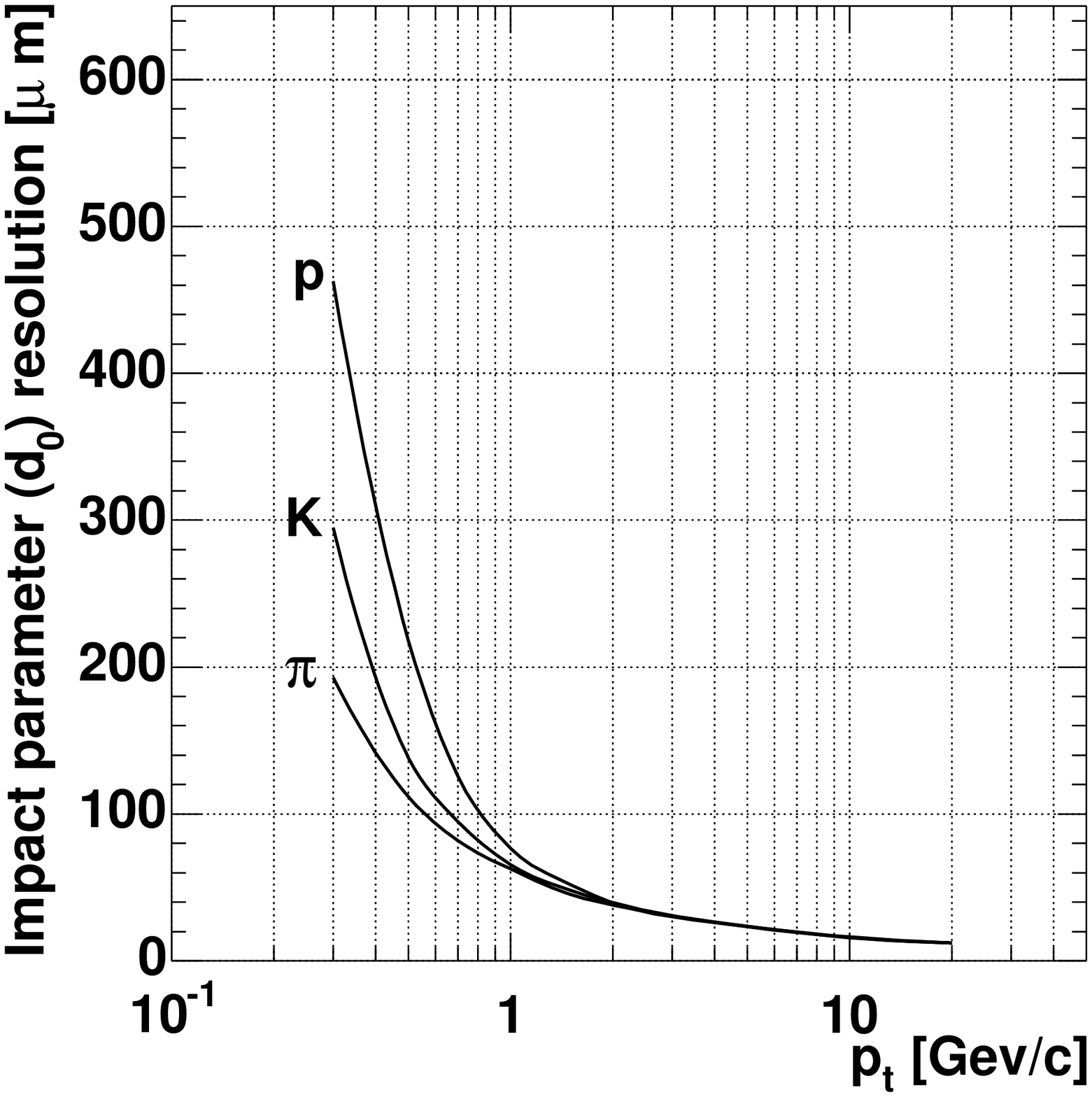}
  \includegraphics[width=0.59\textwidth]{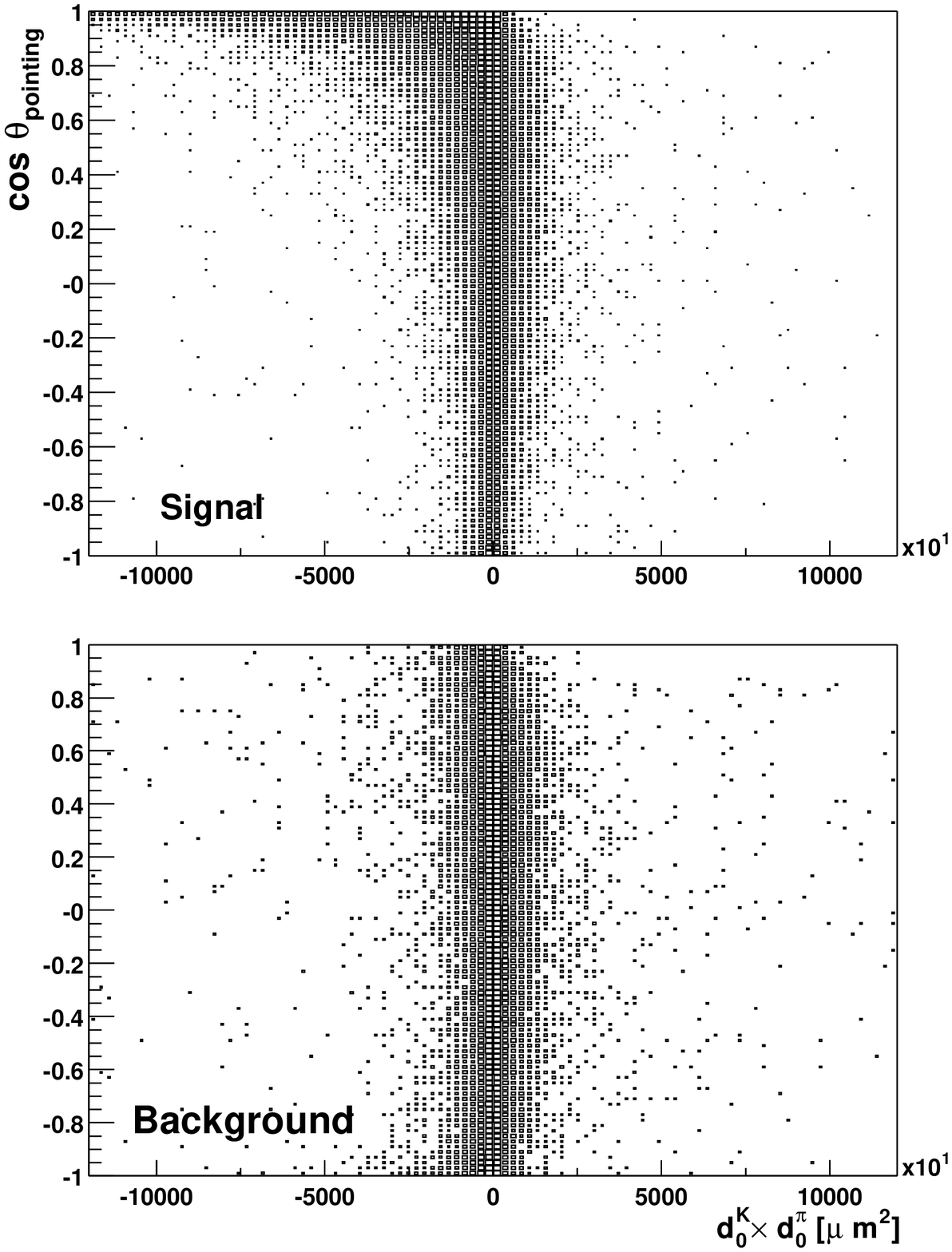}
  \caption{Top: sketch of the $\rm D^0 \to K^-\pi^+$ 
           decay.
           Bottom-left: $\pt$-dependence of the track impact-parameter 
           resolution 
           in central \PbPb~collisions with the ALICE detector,
           for $\pi^\pm$, $\rm K^\pm$ and $\rm p/\overline p$.
           Bottom-right: correlation between the cosine of the pointing 
           angle and the product of the impact parameters for signal 
           and background $\rm D^0\to K^-\pi^+$ candidates.} 
\label{fig:D0combined}
\end{center}
\end{figure}

Figure~\ref{fig:D0combined} (top) shows a sketch of the decay:  
the main feature of this topology is the presence of two tracks with impact 
parameters\footnote{We define as impact parameter, $d_0$, the distance of 
closest approach of the track to the 
interaction vertex, in the plane transverse to the beam direction 
(see sketch in Fig.~\ref{fig:D0combined}).} 
$d_0\simeq 100~\mum$ (the mean 
proper decay length of $\rm D^0$ mesons is $c\tau\simeq 124~\mum$).
Excellent tracking and vertexing capabilities
are necessary to extract the signal out of the huge combinatorial 
background in central \PbPb~collisions, where up to several thousand charged 
particles might be produced per unit of rapidity (simulations were performed 
with $\d N_{\rm charged}/\d y=6000$).

The barrel tracking system of ALICE, 
composed of the Inner Tracking System (ITS), 
the Time Projection Chamber (TPC) and the Transition Radiation Detector (TRD),
embedded in a magnetic field of $0.5$~T, allows track reconstruction in 
the pseudorapidity range $-0.9<\eta<0.9$ with a momentum resolution better than
2\% for $\pt<10~\gev/c$ and an impact-parameter resolution (shown in 
the bottom-left panel of Fig.~\ref{fig:D0combined}) better than 
$60~\mum$ for $\pt>1~\gev/c$, mainly provided by the two layers,
at $r=4$ and $7~\cm$, of silicon pixel detectors of the ITS.

The detection strategy~\cite{D0jpg} to cope with the large combinatorial 
background from the underlying event is based on the selection of 
displaced-vertex topologies. The impact parameter $d_0$ is given a sign 
according to the position of the track with respect to the main interaction 
vertex, so that well-separated signal topologies have impact parameters, 
$d_0^{\rm K}$ and $d_0^\pi$, large and with opposite signs. Therefore, 
the product 
of the impact parameters is required to be negative and large in absolute 
value, e.g. $d_0^{\rm K}\times d_0^\pi<-2\times 10^4~\mum^2$. Another suitable
variable is the pointing angle $\theta_{\rm pointing}$ 
between the reconstructed $\rm D^0$ momentum and its
flight-line (see sketch in Fig.~\ref{fig:D0combined}), 
which is required to be close to zero, e.g. $\cos\theta_{\rm pointing}>0.98$. 
We found that these two variables are strongly correlated for the signal and 
uncorrelated for the background combinations (see bottom-right panel of 
Fig.~\ref{fig:D0combined}); 
therefore, their simultaneous use is extremely efficient in increasing the 
signal-to-background ratio. After such selection, a standard invariant-mass 
analysis can be used to extract the amount of signal. 
The strategy was optimized separately for pp and \PbPb~collisions, as a 
function of the $\rm D^0$ transverse momentum~\cite{thesis}. The requirement 
of $\rm K$ tagging for one of the two tracks 
in the high-resolution Time-Of-Flight (TOF) detector  
and the low value of the magnetic field allow to extend the $\rm D^0$ 
signal extraction down to almost zero transverse momentum.

\begin{figure}[!t]
  \begin{center}
    \includegraphics[width=.45\textwidth]{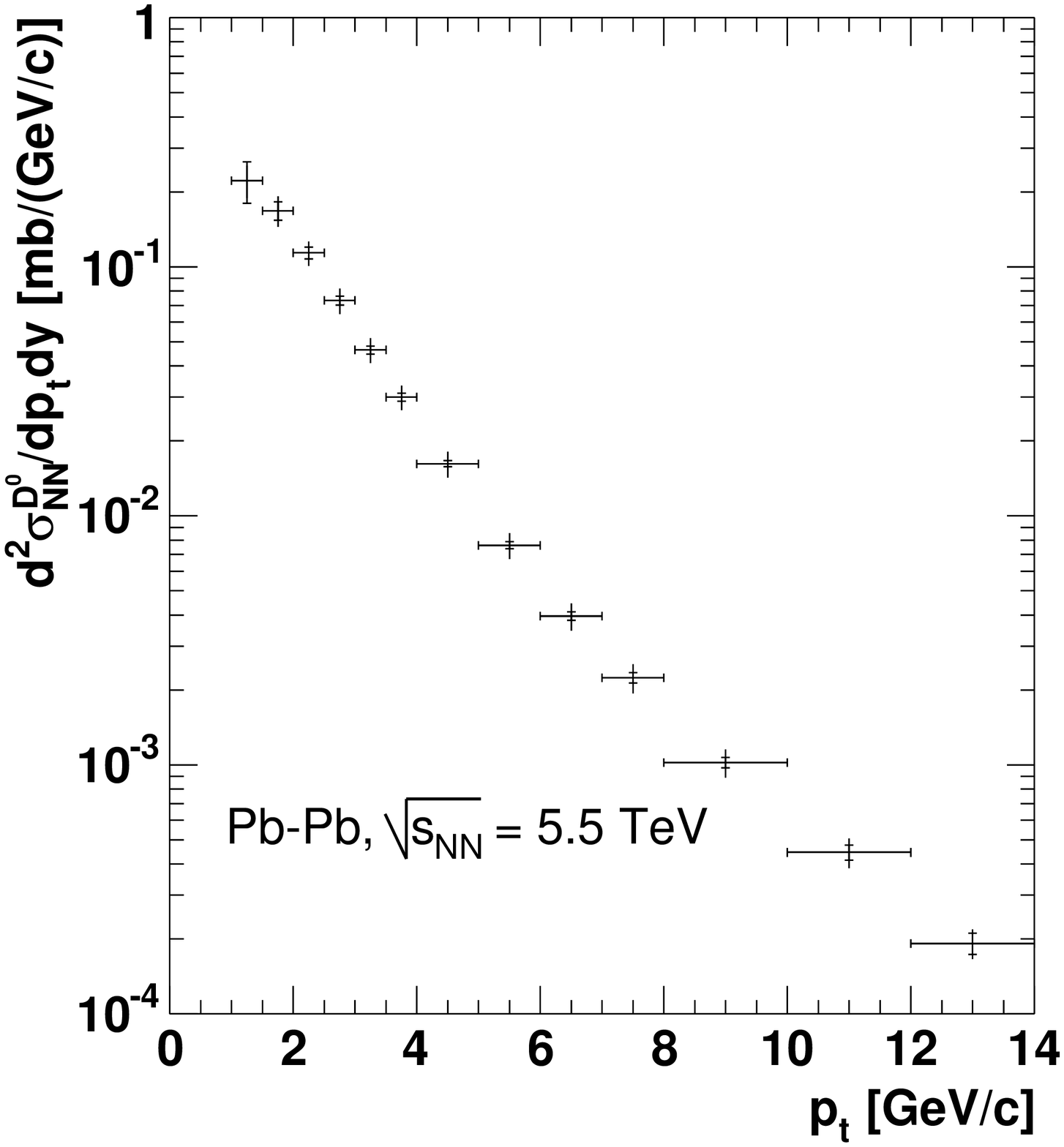}
    \includegraphics[width=.45\textwidth]{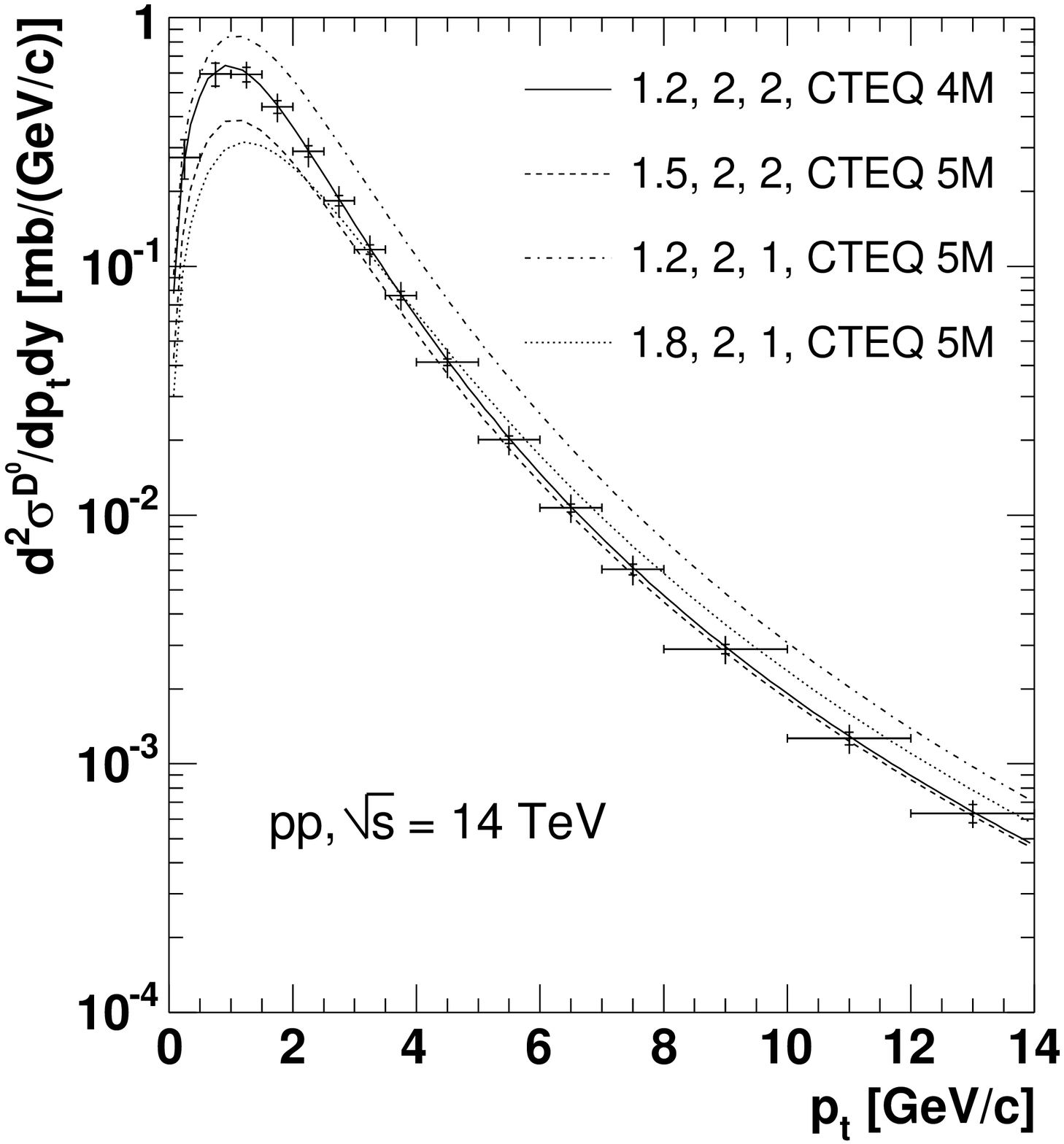}
    \caption{Differential cross section per NN collision 
             for $\Dz$ production as a 
             function of $\pt$, as it can be measured with $10^7$ central  
             \PbPb~events (left) 
             and $10^9$ pp minimum-bias events (right). 
             Statistical (inner bars) and $\pt$-dependent 
             systematic errors (outer bars) are shown. A normalization error
             of 11\% for \PbPb~and 5\% for pp is not shown.
             For pp (right)  
             pQCD predictions obtained with different sets of the input 
             parameters $m_{\rm c}$ [GeV], $\mu_F/\mu_0$, $\mu_R/\mu_0$ 
             ($\mu_0$ is defined in the text) and PDF set are also reported.} 
    \label{fig:D0pt}
  \end{center}
\end{figure}

The expected performance for central \PbPb~(5\% $\sigma^{\rm tot}$) 
and for pp collisions is summarized in Fig.~\ref{fig:D0pt}. The accessible 
$\pt$ range is $1$--$14~\gev/c$ for \PbPb~and $0.5$--$14~\gev/c$ for pp.
The statistical error corresponding to 1 month of \PbPb~data-taking 
($\sim 10^7$ central events) and 9 months of pp data-taking ($\sim 10^9$ 
events) is better than 15--20\% and the systematic error 
(acceptance and efficiency corrections, subtraction of the feed-down from 
${\rm B}\to \Dz+X$ decays, cross-section normalization, 
centrality selection for \PbPb) is better than 20\%~\cite{thesis}.

On the right-hand panel of Fig.~\ref{fig:D0pt} the expected
sensitivity of ALICE for the measurement of the $\Dz$ $\pt$-differential 
cross section is compared to the uncertainty of pQCD 
calculations that we mentioned in Section~\ref{probes}. The 
$\d^2\sigma^{\rm D^0}/\d\pt\d y$ curves 
shown in the figure were obtained by applying the PYTHIA~\cite{pythia} 
fragmentation model to c-quark $\pt$ distributions calculated at NLO 
with the HVQMNR program~\cite{hvqmnr}. The values of $m_{\rm c}$, 
$\mu_F/\mu_0$ and $\mu_R/\mu_0$
were varied similarly to what done by M.~Mangano in Ref.~\cite{yrhvq}.
The errors correspond to the curve obtained with the set of parameters
used for our baseline cross section: 
$m_{\rm c}=1.2~\gev$, $\mu_F/\mu_0=\mu_R/\mu_0=2$ and CTEQ 4M 
PDFs. The figure shows that the broad ALICE $\pt$ coverage, from almost 
zero to about $14~\gev/c$, gives a good constraining
power with respect to the pQCD input parameters. 

\section{ALICE sensitivity to D-meson suppression}
\label{Dquenching}

\begin{figure}[!t]
  \begin{center}
    \includegraphics[width=.62\textwidth]{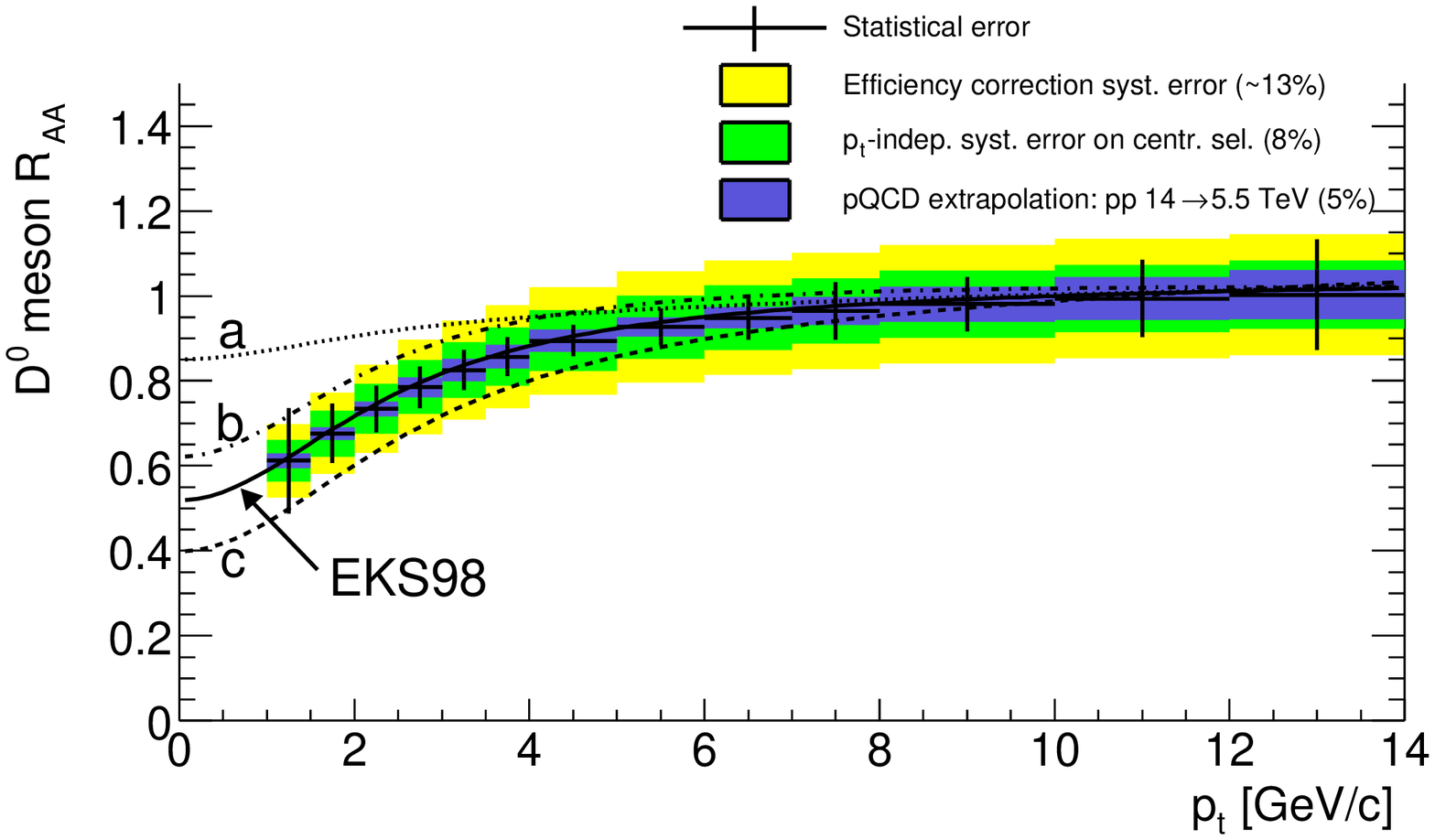}
    \includegraphics[width=.32\textwidth]{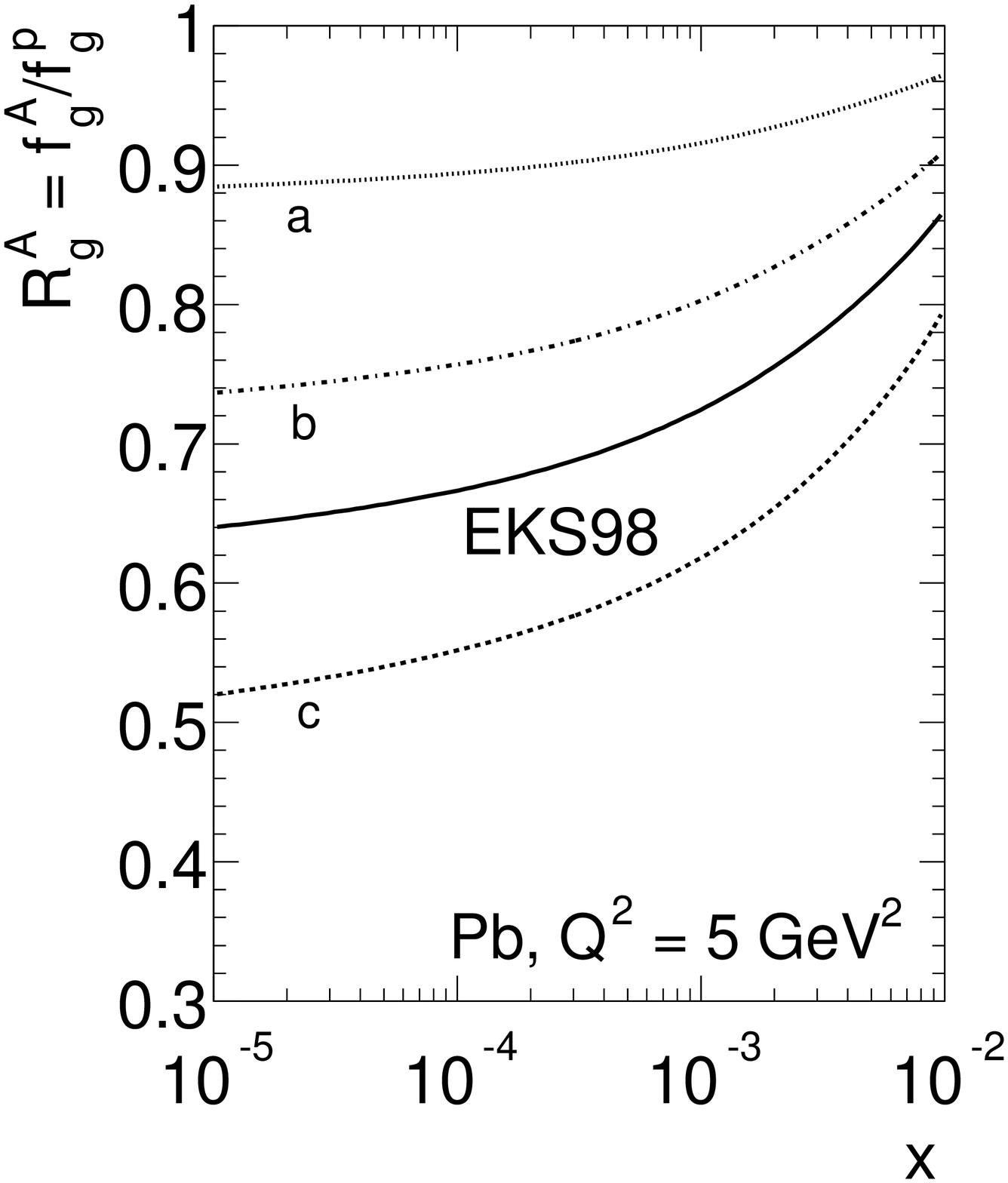}
    \includegraphics[width=0.65\textwidth]{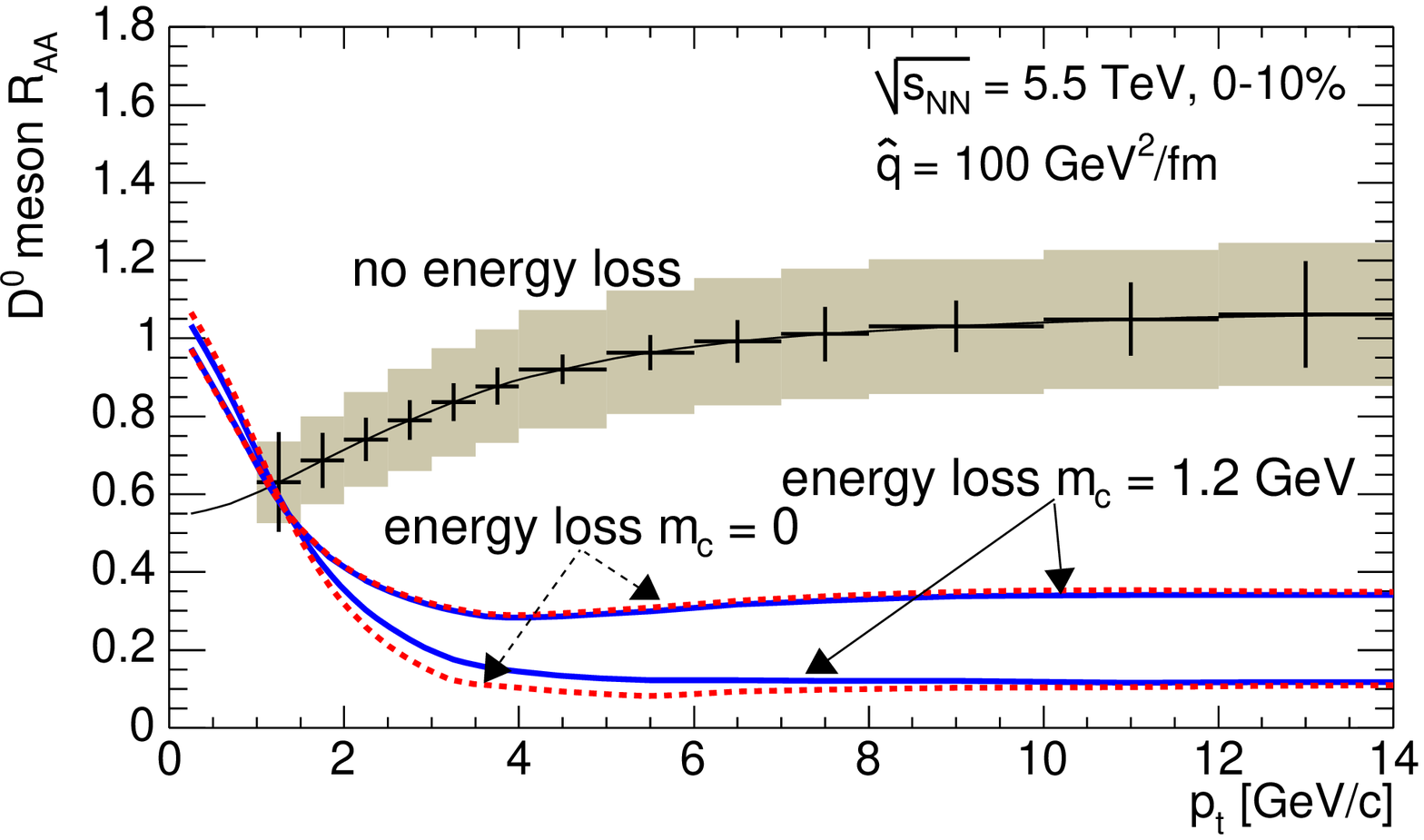}
    \caption{Top-left: $\RAA$ of $\Dz$ mesons (only shadowing included)
             with the statistical errors and the different contributions 
             to the systematic error.
             Top-right: different nuclear modifications of the gluon PDF
             at the scale $Q^2=(2\,m_{\rm c})^2=5~\gev^2$.
             Bottom: $\RAA$ of $\Dz$ mesons, without energy loss
             (`data' points),
             with energy loss for massless 
             quarks ($m_{\rm c}=0$; dashed line) and with energy loss for 
             massive quarks ($m_{\rm c}=1.2~\gev$; solid line).
             The reported errors are: bars = statistical, 
             shaded area = systematic contributions combined.} 
    \label{fig:RAA}
  \end{center}
\end{figure}

The D-meson nuclear modification factor $R_{\rm AA}^{\rm D}(\pt)$, defined 
in Eq.~(\ref{eq:RAAD}), is reported in 
Fig.~\ref{fig:RAA} (top-left panel). Only nuclear shadowing is 
included (no energy loss).
The reported errors are obtained combining 
the previously-mentioned errors in \PbPb~and in pp collisions and 
considering that several systematic 
contributions will partially cancel out in the ratio.
The uncertainty of about 5\% introduced in the
extrapolation of the pp results from 14~TeV to 5.5~TeV by pQCD, estimated
in Ref.~\cite{thesis}, is also shown. 

The effect of shadowing, introduced via the EKS98 parameterization~\cite{eks},
is visible as a suppression at low transverse momenta, $\pt\lsim 7~\gev/c$, 
corresponding to small $x$ ($\lsim 10^{-3}$). 
Since there is a significant uncertainty on the magnitude of shadowing in 
this $x$ region, we studied the effect of such
uncertainty on $\RAA$ by varying the modification of the PDFs in a 
Pb nucleus (shown for gluons in top-right panel of Fig.~\ref{fig:RAA}).
Even in the case of shadowing 50\% stronger than in EKS98 (curves labelled 
``c''), we find $\RAA>0.93$ for $\pt>7~\gev/c$. 
We can, thus, conclude that c-quark energy loss can be cleanly 
studied, being the only expected effect, in the region 
$7\lsim\pt\lsim 15~\gev/c$, where ALICE has a good sensitivity.

In the bottom panel of Fig.~\ref{fig:RAA} we report a recent 
estimate of the suppression of the D-meson nuclear modification factor 
in central \PbPb~collisions 
at the LHC~\cite{new}, obtained 
in the framework of the PQM model~\cite{pqm}, where energy loss 
is simulated in a parton-by-parton approach combining  
the BDMPS `quenching weights'~\cite{qw} and 
a Glauber-model-based definition of the in-medium parton path length.
The quenching weights were specifically calculated for massive partons,
using the formalism developed in Ref.~\cite{armestomassive}. 
The medium transport coefficient at the LHC was set to the value
$\hat{q}=100~\gev^2/\fm$, estimated on the basis of the analysis of 
RHIC data performed in Ref.~\cite{pqm}. The results are plotted as a band
that represents the theoretical uncertainty arising from the finite-energy 
constraint discussed in Section~\ref{probes}. According to this estimate
the c-quark mass does not influence the observed suppression of D mesons
in any significant way.

\section{Conclusions}

We have shown that the direct $\Dz$-meson reconstruction with  
ALICE will allow to measure the charm cross section in pp collisions
and its medium-induced suppression in \PbPb~collisions, thus 
providing stringent experimental constraints to the current 
theoretical understanding both in the domain of perturbative QCD 
calculations and in that of many-body high-density QCD, where 
parton energy loss is computed. 



\bigskip

{\small Discussions with F.~Antinori, E.~Quercigh and
K.~$\check{\mathrm{S}}$afa$\check{\mathrm{r}}$\'\i k are 
gratefully acknowledged. 
The author thanks the organizing committee of the conference 
``Physics at LHC 2004'', where this talk was presented, 
for supporting his participation.}

\bigskip




\begin{thebibliography}{9}

\bibitem{alice}
  ALICE Physics Performance Report, Volume I, CERN/LHCC 2003-049 (2003);\\
  C.~Fabjan, {\it these proceedings}.

\bibitem{muller} 
  B.~Muller, {\it these proceedings}.

\bibitem{armesto} 
  N.~Armesto, {\it these proceedings}.

\bibitem{yrhvq} 
  M.~Bedjidian {\it et al.}, CERN Yellow Report {\it in press},
  arXiv:hep-ph/0311048.

\bibitem{notehvq}
  N.~Carrer and A.~Dainese, ALICE-INT-2003-019 (2003), arXiv:hep-ph/0311225. 

\bibitem{hvqmnr} 
  M.~Mangano, P.~Nason and G.~Ridolfi, Nucl.~Phys.~B~{\bf 373} (1992) 295.

\bibitem{cteq4}
  H.L.~Lai~{\it et al.}, CTEQ Coll., Phys.~Rev.~D~{\bf 55} (1997) 1280. 

\bibitem{eks} 
  K.J.~Eskola, V.J.~Kolhinen and C.A.~Salgado, Eur.~Phys.~J.~C~{\bf 9} 
  (1999) 61.

\bibitem{harris} 
  J.~Harris, {\it these proceedings}.

\bibitem{gyulassywang}
  M.~Gyulassy and X.N.~Wang, Nucl.~Phys.~B~{\bf 420} (1994) 583.

\bibitem{bdmps}
  R.~Baier, Yu.L.~Dokshitzer, A.H.~Mueller, S.~Peign\'e and D.~Schiff, 
  Nucl.~Phys.~B~{\bf 483} (1997) 291; 
  {\em ibidem}~B~{\bf 484} (1997) 265.

\bibitem{wiedemann}
  U.A.~Wiedemann, Nucl.~Phys.~B~{\bf 588} (2000) 303.

\bibitem{glv}
  M.~Gyulassy, P.~L\'evai and I.~Vitev, Nucl.~Phys.~B~{\bf 571} (2000) 197;
  Phys. Rev. Lett. {\bf 85} (2000) 5535;
  Nucl. Phys. B {\bf 594} (2001) 371.

\bibitem{pqm}
  A.~Dainese, C.~Loizides and G.~Paic, arXiv:hep-ph/0406201.  

\bibitem{dokshitzerdeadcone}
  Yu.L.~Dokshitzer, V.A.~Khoze and S.I.~Troyan, J.~Phys.~G~{\bf 17} (1991) 
  1602.

\bibitem{dokshitzerkharzeev}
  Yu.L.~Dokshitzer and D.E.~Kharzeev, Phys.~Lett.~B~{\bf 519} (2001) 199.

\bibitem{armestomassive} 
  N.~Armesto, C.A.~Salgado and U.A.~Wiedemann, Phys.~Rev.~D~{\bf 69} (2004) 
  114003.

\bibitem{D0jpg}
  N.~Carrer, A.~Dainese and R.~Turrisi, J.~Phys.~G~{\bf 29} (2003) 575.

\bibitem{thesis}
  A.~Dainese, Ph.D. Thesis, arXiv:nucl-ex/0311004.

\bibitem{pythia} 
  T.~Sj\"ostrand {\it et al.}, Computer Phys. Commun. {\bf 135} (2001) 238.


\bibitem{new}
  N.~Armesto, A.~Dainese, C.A.~Salgado and U.A.~Wiedemann, 
  {\it in preparation}.

\bibitem{qw}
  C.A.~Salgado and U.A. Wiedemann, Phys.~Rev.~D~{\bf 68} (2003) 014008.

\end{thebibliography}
\end{document}